\documentclass[aps,reprint,superscriptaddress,nobibnotes]{revtex4-1}

\usepackage{siunitx}
\usepackage{graphicx}
\usepackage{hyperref}
\usepackage{amsmath}
\usepackage{bm}

\usepackage{xr}
\usepackage{makecell}
\usepackage{multirow}

\begin{document}

\title{Calculation of intrinsic spin Hall conductivity 
	by Wannier interpolation}

\author{Junfeng~Qiao}
\affiliation{Fert Beijing Institute, BDBC, Beihang University, 
Beijing 100191, China}
\affiliation{School of Electronic and Information Engineering, Beihang 
University, Beijing 100191, China}

\author{Jiaqi~Zhou}
\affiliation{Fert Beijing Institute, BDBC, Beihang University, 
	Beijing 100191, China}
\affiliation{School of Electronic and Information Engineering, Beihang 
	University, Beijing 100191, China}

\author{Zhe~Yuan}
\affiliation{The Center for Advanced Quantum Studies and 
Department of Physics, Beijing Normal University, 100875 Beijing, China}

\author{Weisheng~Zhao}
\email{weisheng.zhao@buaa.edu.cn}
\affiliation{Fert Beijing Institute, BDBC, Beihang University, 
Beijing 100191, China}
\affiliation{School of Electronic and Information Engineering, Beihang 
University, Beijing 100191, China}

\date{\today}

\begin{abstract}
\textit{Ab-initio} calculation of intrinsic spin Hall conductivity  
(SHC) generally requires a strict convergence criterion 
and a dense k-point mesh to sample the Brillouin zone, 
making its convergence challenging and time-consuming. 
Here we present a scheme for efficiently and accurately 
calculating SHC based on maximally 
localized Wannier function (MLWF). The quantities needed by the 
Kubo formula of SHC are derived in the space of MLWF and it 
is shown that only the Hamiltonian, the overlap and the 
spin operator matrices are required from the initial 
\textit{ab-initio} calculation. The computation of these matrices 
and the interpolation of Kubo formula on a dense k-point mesh 
can be easily achieved. 
We validate our results by prototypical calculations on 
fcc Pt and GaAs, which demonstrate that the Wannier 
interpolation approach is of high accuracy and  
efficiency. Calculations of $\alpha$-Ta and $\beta$-Ta 
show that SHC of $\beta$-Ta is 2.7 times of  
$\alpha$-Ta, while both have the opposite sign relative 
to fcc Pt and are an order of magnitude smaller than Pt. 
The calculated spin Hall angle of $-0.156$, is quite 
consistent with previous experiment on $\beta$-Ta, further 
suggesting intrinsic contribution may dominate in 
$\beta$-Ta.  
Our approach could facilitate large-scale SHC calculations, 
and may benefit the discovery of materials with high intrinsic SHC.

\end{abstract}

\maketitle

\section{Introduction}

The spin Hall effect (SHE) is the phenomenon in which transverse 
pure spin current can be generated by applying an electric field 
\cite{Sinova2015}.
In recent years, the utilization of SHE for the magnetization 
switching of magnetic tunnel junction has attracted lots 
of interest due to its potentially 
fast switching speed and low power consumption 
\cite{Liu2012,Liu2012a}. SHE, together with other mechanisms 
like Rashba and Dresselhaus effects, give rise to 
the spin orbit torque (SOT) \cite{Garello2013,Manchon2009Mar,Sato2018Sep}, 
which provides an alternative switching mechanism apart 
from spin transfer torque. 
The SHE can be separated into intrinsic SHE 
which is directly derived from relativistic band 
structure \cite{Murakami1348,Sinova2004a}, and 
extrinsic side-jump and skew-scattering SHE 
which are related to scattering \cite{Wang2016a}. 
The intrinsic SHE, significantly contributing 
to the total SHE in materials with 
strongly spin-orbit-coupled bands \cite{Sinova2015}, 
can be calculated accurately based on \textit{ab-initio} 
theories. According to the Kubo formula, 
the spin Hall 
conductivity (SHC) can be written as \cite{Yao2005,Gradhand2012,Matthes2016} 
\begin{align}
\begin{split}
\label{equ:kubo_shc}
\sigma_{xy}^{\text{spin}z}(\omega) & = 
\hbar \int_{\text{BZ}} \frac{d^3 k}{(2\pi)^3}
\sum_{n} f_{n\bm{k}} \\
& \times \sum_{m \neq n}
\frac{2\operatorname{Im}[\langle n\bm{k}| 
	\hat{j}_{x}^{\text{spin}z}|m\bm{k}\rangle
	\langle m\bm{k}| -e\hat{v}_{y}|n\bm{k}\rangle]}
{(\epsilon_{n\bm{k}}-\epsilon_{m\bm{k}})^2-(\hbar\omega +i\eta)^2},
\end{split}
\end{align}
where $n,m$ are band indexes, $\epsilon_n,\epsilon_m$ 
are the eigenvalues, $f_{n\bm{k}}$ is the Fermi 
distribution function, BZ is the 
first Brillouin zone, 
$\hat{j}_{x}^{\text{spin} z} = 
\frac{1}{2}\{\hat{s}_z,\hat{v}_x\}$ is the spin current operator and $\hat{s}_z=\frac{\hbar}{2}\hat{\sigma}_z$ is the spin operator, 
$\hat{v}_{y} = \frac{1}{\hbar} 
\frac{\partial{H(\bm{k})}}{\partial{k_y}}$ is the 
velocity operator and the frequency $\omega$ 
and $\eta$ are set to zero in the direct current (dc) 
clean limit. 

The calculation of SHC can be performed by 
direct evaluation of Equ.
(\ref{equ:kubo_shc}) or its equivalent form on 
the \textit{ab-initio} wave function, which 
could be acquired by plane wave (PW) or 
all-electron linear muffin-tin orbital method (LMTO) 
\cite{Yao2004,Guo2005,Guo2008}. 
However, these kinds of direct 
evaluations are very time-demanding, usually 
an extremely dense k-point mesh (k-mesh) on the 
order of one million is needed
\cite{Guo2008}. Another available approach is 
evaluating Equ.(\ref{equ:kubo_shc}) on a set 
of tight-binding parameters \cite{Sun_2016}. 
Efficient as well as fully 
\textit{ab-initio} evaluation of Equ.(\ref{equ:kubo_shc}) 
is in urgent need.
 
To mitigate this problem, the direct evaluation of Kubo formula 
should be avoided. Anomalous Hall effect (AHE) is a phenomenon closely 
related to SHE, and a very successful scheme for the 
evaluation 
of anomalous Hall conductivity (AHC) has been developed 
\cite{Wang2006} based on 
maximally localized Wannier function (MLWF), which utilizes 
the remaining gauge freedom of the Bloch function 
\cite{Marzari1997,Souza2001}. With an unitary 
gauge transformation, the discontinuous Bloch gauge is transformed 
into a smooth Wannier gauge. Followed by a Fourier transformation, 
the real space Hamiltonian and other quantities can be constructed, 
which are localized in real space because of the smoothness of 
Wannier gauge. The maximally localized real space quantities thus 
enable the interpolation on arbitrary dense k-mesh. 
The Wannier approach 
is both efficient and accurate, since in essence MLWF can 
be viewed as a tight-binding basis while at the same time preserves 
the accuracy of \textit{ab-initio} calculation in 
the energy window of interest, due to its real space maximally 
localization. 
Inspired by the method for AHC, we derived the 
formulas for the evaluation of SHC based on Wannier interpolation. 

The Wannier interpolation approach for calculating SHC is composed of 
two parts: the construction of MLWF from \textit{ab-initio} 
wave function and the calculation of SHC based on MLWF. Generally, only 
a relatively coarse k-mesh \textit{ab-initio} calculation is enough 
for the construction of MLWF. Afterwards, the interpolation of needed 
quantities on a dense k-mesh can be easily achieved 
with moderate computational burden. 

The paper is organized as follows. In Sec.\ref{sec:bg} we describe 
the basic theories and introduce 
the essential quantities we need to evaluate in our Wannier-based 
approach. 
In Sec.\ref{sec:deriv} we provide the detailed derivations on 
the MLWF basis. To 
validate our approach, exemplary calculations of Pt and GaAs 
are shown in Sec.\ref{sec:pt} and \ref{sec:gaas}.  
Sec.\ref{sec:ta} describes SHC calculations of 
$\alpha$-Ta and $\beta$-Ta. Finally, Sec.\ref{sec:sum} 
contains a brief summary.

\section{Definitions and Background\label{sec:bg}}
First we give a basic description of Kubo formula 
for SHC. Since the Wannier interpolation approach for AHC has 
been well established, a comparison of SHC and AHC is beneficial.
Then we provide a brief and self-contained 
introduction to the MLWF. More details on MLWF can be 
found in Ref. \cite{Marzari1997,Souza2001,Marzari2012}. 

\subsection{Kubo formula}
The general form of Kubo formula for AHC and SHC is given by 
\cite{Yao2004,Guo2005,Matthes2016} 
\begin{align}
\begin{split}
\label{equ:kubo}
\sigma_{xy}(\omega) = & \frac{\hbar}{V N_k^3}
\sum_{\bm{k}}\sum_{n} f_{n\bm{k}} \\
& \times \sum_{m \neq n}
\frac{2\operatorname{Im}[\langle n\bm{k}| \hat{j}_{x}|m\bm{k}\rangle
	\langle m\bm{k}| -e\hat{v}_{y}|n\bm{k}\rangle]}
{(\epsilon_{n\bm{k}}-\epsilon_{m\bm{k}})^2-(\hbar\omega +i\eta)^2},
\end{split}
\end{align}
where $\hat{j}_{x}=-e\hat{v}_{x},
\frac{1}{2}\{\hat{s}_z,\hat{v}_x\}$ 
for AHC and SHC, 
respectively. Here the integral in Equ.(\ref{equ:kubo_shc}) 
is replaced by numerical sum, and $V$ is the primitive cell 
volume, $N_k^3$ is the number of k-points in the BZ. 
The SHC is multiplied by 
$\frac{-2e}{\hbar}$ to convert it into the unit of 
\si{S/length}, the same as that of AHC. 
Comparing the Kubo formulas for AHC and SHC, 
the only difference is the spin current operator 
matrix elements 
$\langle n\bm{k}| \hat{j}_{x}^{\text{spin} z}|m\bm{k}\rangle$, 
which is the key quantity we need to 
evaluate. 

To facilitate further analysis, we separate the 
Equ.(\ref{equ:kubo}) into the band-projected 
Berry curvature-like term
\begin{align}
\begin{split}
\label{equ:kubo_shc_berry}
\Omega_{n,xy}^{\text{spin}z}(\bm{k}) = {\hbar}^2 \sum_{
	m\ne n}\frac{-2\operatorname{Im}[\langle n\bm{k}| 
	\frac{1}{2}\{\hat{\sigma}_z,\hat{v}_x\}|m\bm{k}\rangle
	\langle m\bm{k}| \hat{v}_{y}|n\bm{k}\rangle]}
{(\epsilon_{n\bm{k}}-\epsilon_{m\bm{k}})^2-(\hbar\omega+i\eta)^2}
\end{split}
\end{align}
and the SHC is the sum over occupied bands
\begin{equation}
\label{equ:kubo_shc_sum}
\sigma_{xy}^{\text{spin}z}(\omega) = 
-\frac{e^2}{\hbar}\frac{1}{V N_k^3}\sum_{\bm{k}}\sum_{n}
f_{n\bm{k}} \Omega_{n,xy}^{\text{spin}z}(\bm{k}).
\end{equation}
The unit of the $\Omega_{n,xy}^{\text{spin}z}(\bm{k})$ 
is $\text{length}^{2}$, and the unit of  $\sigma_{xy}^{\text{spin}z}$ 
is $\frac{e^2}{\hbar}\frac{1}{\text{length}}$ 
[note $\frac{e^2}{\hbar} \simeq$ \SI{2.434e-4}{S}]. 
To convert into the unit $\frac{\hbar}{e} {\text{S/length}}$, 
the $\sigma_{xy}^{\text{spin}z}$ should be multiplied 
by $\frac{\hbar}{-2e}$. 
The case of $\omega=0$ corresponds to 
dc SHC while that of $\omega\ne0$ 
corresponds to alternating current (ac) SHC. 
 
\subsection{Wannier interpolation}
\subsubsection{Construction of MLWF}
The Kohn-Sham equation for the periodic part of the Bloch function is 
written as 
\begin{equation}
\hat{H}_{\bm{k}} u_{n\bm{k}} = \epsilon_{n\bm{k}} u_{n\bm{k}},
\end{equation}
where $\bm{k}$ is the k-point vector, 
$n$ is the band index, 
$u_{n\bm{k}}(\bm{r}) = e^{-i \bm{k} \bm{r}} \psi_{n\bm{k}}(\bm{r})$ 
is the periodic part of the Bloch function 
$\psi_{n\bm{k}}(\bm{r})$. 
$\hat{H}_{\bm{k}}$ is the transformed Hamiltonian 
$\hat{H}_{\bm{k}} = e^{-i \bm{k} \bm{r}} \hat{H} e^{i \bm{k} \bm{r}}$ 
and $\epsilon_{n\bm{k}}$ is the eigenvalue. 

Usually one needs a lot of plane waves to expand 
the $u_{n\bm{k}}$ in PW method and the diagonalization of 
the Hamiltonian matrix is performed on 
each k-point. An alternative representation is the real space 
Wannier function $|\bm{R}n\rangle$, which can be viewed 
as the Fourier transform of the Bloch wave function 
\begin{equation}
\label{equ:wf_ft}
| \bm{R} n \rangle = \frac{V}{(2 \pi)^3} \int_{\text{BZ}} d \bm{k}
e^{-i \bm{k} \cdot \bm{R}} | \psi_{n \bm{k}} \rangle,
\end{equation}
where $V$ is the volume of the real-space primitive cell 
and BZ is the Brillouin zone.

In principle, a smooth function in real space results 
in a localized function in its reciprocal space, and vice versa. 
However, it is not naturally guaranteed that the simply summed 
Bloch function of Equ. (\ref{equ:wf_ft}) results in a 
smooth function $|\bm{R} n \rangle$ in real space. 
Fortunate enough, there is a gauge freedom left in the 
definition of Bloch function, we can replace 
$|\psi_{n\bm{k}} \rangle$ by 
\begin{equation}
| \tilde{\psi}_{n\bm{k}} \rangle = e^{i \varphi_n(\bm{k})} 
| \psi_{n\bm{k}} \rangle,
\end{equation}
or equivalently, 
\begin{equation}
| \tilde{u}_{n\bm{k}} \rangle = e^{i \varphi_n(\bm{k})} 
| u_{n\bm{k}} \rangle
\end{equation}
without changing the physical description of the system. 
The $\varphi_n(\bm{k})$ can be any real function that is 
periodic in reciprocal space \cite{Marzari2012}. 
We can utilize this freedom to construct localized WFs 
in real space, the so-called maximally localized Wannier 
function.

We define the $J$ dimensional unitary transformation 
$\mathcal{U}(\bm{k})$ which takes the 
original Bloch function $| u_{n\bm{k}}^{(0)} \rangle$ to 
the smoothed function $| \tilde{u}_{n \bm{k}} \rangle$ as
\begin{equation}
\label{equ:u2w}
| \tilde{u}_{n\bm{k}} \rangle = \sum_{m=1}^{J} | 
u_{m\bm{k}}^{(0)} \rangle \mathcal{U}_{mn}(\bm{k}),
\end{equation}
where $J$ is the number of states need to be considered 
for our targeted physical properties. 
We call this unitary transformation as the 
transformation from Bloch gauge to Wannier gauge. 
Thus, the Fourier transformation pair between the 
smoothed Bloch functions and the MLWFs are 
\begin{equation}
\begin{split}
| \bm{R} n \rangle & = \frac{1}{N_k^3} \sum_{\bm{k}}
e^{-i \bm{k} \cdot \bm{R}} | \tilde{u}_{n \bm{k}} \rangle, \\
& \Updownarrow \\
| \tilde{u}_{n \bm{k}} \rangle & = \sum_{\bm{R}}
e^{i \bm{k} \cdot \bm{R}} | \bm{R} n \rangle,
\end{split}
\end{equation}
where $N_k^3$ is the number of points in BZ.

The above mentioned procedure is suitable for 
a group of isolated bands \cite{Marzari1997}, 
e.g. the occupied valence 
bands of an insulator. However, for entangled bands, 
a process called disentanglement should be adopted \cite{Souza2001}, 
in which $J$ smoothly varying $|\tilde{u}_{n \bm{k}} \rangle$ 
are extracted from $\mathcal{J}_{\bm{k}} > J$ original 
Bloch bands, where $\mathcal{J}_{\bm{k}}$ can be varied 
throughout the BZ. Using a set of $J$ localized 
trial orbitals $g_n(r)$ and projecting them onto 
the original Bloch states
\begin{equation}
|\phi_{n\bm{k}}\rangle = \sum_{m=1}^{\mathcal{J}_{\bm{k}}}
|\psi_{m\bm{k}}^{(0)}\rangle \langle\psi_{m\bm{k}}^{(0)} | g_n \rangle,
\end{equation}
after normalization 
\begin{equation}
|\psi_{m\bm{k}}\rangle = \sum_{m=1}^{J}
|\phi_{m\bm{k}}\rangle (S_{\bm{k}}^{-1/2})_{mn}
\end{equation}
we acquire a set of $J$ smooth Bloch-like states. The 
overlap matrix 
$(A_{\bm{k}})_{mn} = \langle \psi_{m\bm{k}}|g_n\rangle$ 
is of dimension $\mathcal{J}_{\bm{k}}\times J$ and 
$(S_{\bm{k}})_{mn} = \langle \phi_{m\bm{k}}|\phi_{n\bm{k}}
\rangle _{V} = (A_{\bm{k}}^+ A_{\bm{k}})_{mn}$ where 
$A_{\bm{k}}^+$ is the conjugate transpose of $A_{\bm{k}}$. 
Other than projecting onto the trial orbitals $g_n(r)$, 
we can adopt an iterative method to obtain an optimally 
smooth space of $J$ Bloch-like states at each $\bm{k}$ 
which are the linear combinations of the original 
$\mathcal{J}_{\bm{k}}$ states. 

In summary, by disentanglement we construct an 
optimally smooth space of $J$ Bloch-like states from 
initial $\mathcal{J}_{\bm{k}}$ Bloch states. Then 
by a gauge-selection step we obtain $J$ individually 
smooth Wannier gauge states. From now on we write the 
smooth varying $| \tilde{u}_{n \bm{q}} \rangle$ as 
$| u_{n\bm{q}}^{(W)} \rangle$ to be consistent with 
latter derivations, and we use $\bm{q}$ rather than 
$\bm{k}$ to represent the k-point of the initial 
\textit{ab-initio} k-mesh, while the $\bm{k}$ symbol is 
reserved for the Wannier interpolation k-mesh used 
in the following steps. 
The combined result is 
\begin{equation}
|\psi_{n\bm{q}}^{(W)}\rangle = 
\sum_{m=1}^{\mathcal{J}_{\bm{q}}} 
|\psi_{m\bm{q}}^{(0)}\rangle V_{\bm{q},mn}.
\end{equation}
Here, $V_{\bm{q},mn}$ is a $\mathcal{J}_{\bm{k}}\times J$ 
dimensional matrix.

\subsubsection{Wannier interpolation of Hamiltonian matrix}
After MLWF has been constructed, we interpolate 
operators on a dense k-mesh to get a converged 
result. In this section we use the interpolation 
of the Hamiltonian matrix $H_{nm}(\bm{k}) = 
\langle \psi_{n\bm{k}}| \hat{H}|
\psi_{m\bm{k}}\rangle$ as an example to illustrate 
the key ideas behind Wannier interpolation.

For the reciprocal space Hamiltonian operator 
$\hat{\bm{H}}(\bm{q})$, we define the $J \times J$ Hamiltonian 
matrix in the Wannier gauge as 
\begin{equation}
H_{nm}^{(W)}(\bm{q}) = \langle u_{n\bm{q}}^{(W)} | 
\hat{\bm{H}}(\bm{q}) | u_{m\bm{q}}^{(W)} \rangle
= [V^+ (\bm{q}) H^{(0)}(\bm{q}) V(\bm{q})]_{nm},
\end{equation}
where $H_{nm}^{(0)}(\bm{q}) = 
\mathcal{E}^{(0)}_{n \bm{q}} \delta_{nm}$ 
is the diagonal Hamiltonian matrix of the original 
Bloch states, $\delta_{nm}$ is  the 
Kronecker delta function. 
If diagonalizing $H_{nm}^{(W)}(\bm{q})$ by 
\begin{equation}
U^+(\bm{q}) H^{(W)}(\bm{q}) U(\bm{q}) = H^{(H)}(\bm{q}),
\end{equation}
where $H_{nm}^{(H)}(\bm{q}) = \mathcal{E}_{n \bm{q}}^{(H)} \delta_{nm}$, 
then $\mathcal{E}_{n \bm{q}}^{(H)}$ will be identical to the original 
\textit{ab-initio} $\mathcal{E}^{(0)}_{n \bm{q}}$ in the 
range of $n=1,2,...,J$. 
Transforming the Hamiltonian operator from reciprocal space to real space, 
\begin{equation}
H_{nm}^{(W)}(\bm{R}) = \frac{1}{N_{q}^3} \sum_{\bm{q}} e^{-i \bm{q} \cdot \bm{R}}
H_{nm}^{(W)}(\bm{q}),
\end{equation}
and then performing inverse Fourier transform 
\begin{equation}
\label{inv_trans_oper}
H_{nm}^{(W)}(\bm{k}) = \sum_{\bm{R}} e^{i \bm{k} \cdot \bm{R}} 
H_{nm}^{(W)}(\bm{R}),
\end{equation}
we succeed in interpolating the Hamiltonian operator on 
arbitrary k-point $\bm{k}$. 

Since the Wannier functions (WF) we chose are maximally localized, the 
$H_{nm}^{(W)}(\bm{R})$ is expected to be well localized 
in real space, a few $\bm{R}$ are sufficient in the 
sum of Equ. (\ref{inv_trans_oper}). 

The final step is to diagonalize $H_{nm}^{(W)}(\bm{k})$, 
\begin{equation}
U^+(\bm{k}) H^{(W)}(\bm{k}) U(\bm{k}) = H^{(H)}(\bm{k}),
\end{equation}
then the acquired eigenvalues and gauge transformation 
matrix $U(\bm{k})$ on arbitrary k-point $\bm{k}$ 
can be used for latter extractions of the targeted physical properties. 
We comment here that since $H^{(W)}(\bm{k})$ are of dimensions 
$J \times J$, their diagonalizations are very ``cheap'', compared 
with the diagonalizations of the Hamiltonian matrices in PW method. 

Apart from the Hamiltonian matrix $\langle \psi_{m\bm{k}}| 
\hat{H}|\psi_{n \bm{k}}\rangle$, 
another useful quantity for the calculation of SHC is the 
velocity operator matrix 
$\langle \psi_{m\bm{k}}| \hat{v}_{y}|
\psi_{n \bm{k}}\rangle$. 
By using integration by parts, we arrive at 
\begin{align}
\begin{split}
\langle \psi_{m\bm{k}}| \hat{v}_{y}|\psi_{n\bm{k}}\rangle = &
\frac{1}{\hbar}
\langle \psi_{m\bm{k}}| \frac{\partial \hat{H}_{\bm{k}}}
{\partial \bm{k}_{y}}|\psi_{n\bm{k}}\rangle \\
= & \frac{1}{\hbar}\frac{\partial \mathcal{E}_{n\bm{k}}}
{\partial \bm{k}_{y}} \delta_{mn} + \frac{i}{\hbar} 
(\mathcal{E}_{m\bm{k}}-\mathcal{E}_{n\bm{k}})A_{mn,y}(\bm{k}),
\end{split}
\end{align}
\begin{equation}
A_{mn,y}(\bm{k}) = i\langle u_{m\bm{k}}|
\partial_{y} u_{n\bm{k}} \rangle,
\end{equation}
where $\partial_y = \frac{\partial}{\partial k_y}$.
The Wannier interpolation of the 
matrix $A_{mn,y}$ has been developed in the 
calculation of AHC, detailed derivations can be 
found in Ref. \cite{Wang2006}

\section{Derivation of SHC for Wannier interpolation\label{sec:deriv}}

Expanding the spin current operator as 
$\hat{j}_{x}^{\text{spin} z} = \frac{1}{2}
(\hat{s}_z \hat{v}_x + \hat{v}_x \hat{s}_z)$, 
since $\langle \psi_{n\bm{k}}| \hat{s}_z \hat{v}_x |\psi_{m\bm{k}}\rangle^* = 
\langle \psi_{m\bm{k}}| (\hat{s}_z \hat{v}_x)^\dagger |\psi_{n\bm{k}}\rangle = 
\langle \psi_{m\bm{k}}| \hat{v}_x \hat{s}_z |\psi_{n\bm{k}}\rangle$, we define 
\begin{equation}
B_{nm}(\bm{k}) = \langle \psi_{n\bm{k}}|\hat{s}_z \hat{v}_x |\psi_{m\bm{k}}\rangle,
\end{equation}
thus 
\begin{equation}
\label{equ:js}
\langle \psi_{n\bm{k}}|\hat{j}_{x}^{\text{spin} z}|\psi_{m\bm{k}}\rangle 
= \frac{1}{2}[B(\bm{k}) + B^+(\bm{k})]_{nm}.
\end{equation}
Considering the velocity operator  
$\hat{v}_{x} = \frac{1}{\hbar} 
\frac{\partial{H(\bm{k})}}{\partial{k_x}}$ 
and using integration by parts, we arrive at 
\begin{align}
\begin{split}
\label{equ:b3}
B_{nm}(\bm{k}) = & \frac{1}{\hbar}
\frac{\partial \mathcal{E}_{m\bm{k}}}{\partial k_x}
\langle u_{n\bm{k}}| \hat{s}_z | u_{m\bm{k}} \rangle \\
& + \frac{\mathcal{E}_{m\bm{k}}}{\hbar}
\langle u_{n\bm{k}}| \hat{s}_z | \partial_x u_{m\bm{k}} \rangle \\
& - \frac{1}{\hbar}
\langle u_{n\bm{k}}| \hat{s}_z \hat{H}_{\bm{k}} | \partial_x u_{m\bm{k}} \rangle.
\end{split}
\end{align}
For the simplicity of 
latter derivations, we define 
\begin{equation}
\label{equ:snm}
S_{nm}^{(H)}(\bm{k}) = \langle u_{n\bm{k}}^{(H)}| \hat{s}_z | 
u_{m\bm{k}}^{(H)} \rangle,
\end{equation}
\begin{equation}
\label{equ:knm}
K_{nm}^{(H)}(\bm{k}) = \langle u_{n\bm{k}}^{(H)}| \hat{s}_z | 
\partial_x u_{m\bm{k}}^{(H)} \rangle,
\end{equation}
\begin{equation}
\label{equ:lnm}
L_{nm}^{(H)}(\bm{k}) = \langle u_{n\bm{k}}^{(H)}| \hat{s}_z 
\hat{H}_{\bm{k}} | \partial_x u_{m\bm{k}}^{(H)} \rangle,
\end{equation}
where the superscript $(H)$ serves as a reminder that these 
$|u_{n\bm{k}}^{(H)}\rangle$ lie in the Bloch gauge. 
In short,
\begin{equation}
B_{nm}^{(H)}(\bm{k}) = \frac{1}{\hbar}
\frac{\partial \mathcal{E}_{m\bm{k}}}{\partial k_x}
S_{nm}^{(H)}(\bm{k}) + \frac{\mathcal{E}_{m\bm{k}}}{\hbar}
K_{nm}^{(H)}(\bm{k}) - \frac{1}{\hbar}
L_{nm}^{(H)}(\bm{k}).
\end{equation}

Then we need to transform to Wannier gauge by using
\begin{equation}
\label{equ:u2h}
|u_{n\bm{k}}^{(H)}\rangle = \sum_m |u_{m\bm{k}}^{(W)}\rangle U_{mn}.
\end{equation}
Substituting Equ.(\ref{equ:u2h}) into Equ.(\ref{equ:snm}) 
(\ref{equ:knm}) (\ref{equ:lnm}), we arrive at 
\begin{equation}
S^{(H)} = U^+ \langle u^{(W)}| \hat{s}_z | u^{(W)} \rangle U ,
\end{equation}
\begin{align}
\begin{split}
K^{(H)} = & U^+ \langle u^{(W)}| \hat{s}_z | 
\partial_x u^{(W)} \rangle U \\
& + U^+ \langle u^{(W)}| \hat{s}_z | 
u^{(W)} \rangle U D_x^{(H)},
\end{split}
\end{align}
\begin{align}
\begin{split}
L^{(H)} = & U^+ \langle u^{(W)}| \hat{s}_z \hat{H} | 
\partial_x u^{(W)} \rangle U \\
& + U^+ \langle u^{(W)}| \hat{s}_z \hat{H} | 
u^{(W)} \rangle U D_x^{(H)},
\end{split}
\end{align}
where
\begin{equation}
D_x^{(H)} = U^ + \partial_x U,
\end{equation} 
and the $D_x^{(H)}$ has been computed in the 
calculation of velocity operator. 
From now on the subscript $nm$ and the k-point $\bm{k}$ are 
omitted for conciseness and matrix multiplication are implied.

To evaluate the Wannier gauge matrices like 
$\langle u^{(W)}| \hat{s}_z | u^{(W)} \rangle$, we need to 
transform $| u_{n\bm{k}}^{(W)} \rangle$ into its real space representation 
$| \bm{R}n \rangle$  
\begin{equation}
\label{equ:fft}
|\bm{R}n \rangle = \frac{1}{N_k^3} \sum_{\bm{k}} e^{i\bm{k}(\bm{r}-\bm{R})} |u_{n\bm{k}}^{(W)}\rangle
\end{equation}
by its inverse Fourier transform
\begin{equation}
\label{equ:invfft}
|u_{n\bm{k}}^{(W)}\rangle = \sum_{\bm{R}} e^{-i\bm{k}(\bm{r}-\bm{R})} |\bm{R}n\rangle.
\end{equation}
The maximal localization ensures that a minimum set of nearest 
neighbor $\bm{R}$-points is sufficient for the inverse Fourier transform 
Equ.(\ref{equ:invfft}), which enables the accurate interpolation 
of wave function on arbitrary k-point. This is the core function 
of MLWF and accounts for its successful evaluation of many physical 
quantities such as AHC \cite{Wang2006}, 
orbital magnetization \cite{Lopez_2012}, and etc. \cite{Wang2007a}

Substituting Equ.(\ref{equ:invfft}) to the Wannier gauge matrices, 
we arrive at 
\begin{equation}
\label{equ:usu_invfft}
\langle u_{\bm{k}}^{(W)}| \hat{s}_z | u_{\bm{k}}^{(W)} \rangle =
\sum_{\bm{R}} e^{i\bm{k}\bm{R}} \langle \bm{0}|\hat{s}_z|\bm{R}\rangle,
\end{equation}
\begin{equation}
\label{equ:usdu_invfft}
\langle u_{\bm{k}}^{(W)}| \hat{s}_z | \partial_x u_{\bm{k}}^{(W)} \rangle =
-i \sum_{\bm{R}} e^{i\bm{k}\bm{R}} \langle \bm{0}|\hat{s}_z (\bm{r}-\bm{R})_x|\bm{R}\rangle,
\end{equation}
\begin{equation}
\label{equ:ushdu_invfft}
\langle u_{\bm{k}}^{(W)}| \hat{s}_z \hat{H} | \partial_x u_{\bm{k}}^{(W)} \rangle =
-i \sum_{\bm{R}} e^{i\bm{k}\bm{R}} \langle \bm{0}|\hat{s}_z \hat{H} (\bm{r}-\bm{R})_x|\bm{R}\rangle,
\end{equation}
\begin{equation}
\label{equ:ushu_invfft}
\langle u_{\bm{k}}^{(W)}| \hat{s}_z \hat{H} | u_{\bm{k}}^{(W)} \rangle =
\sum_{\bm{R}} e^{i\bm{k}\bm{R}} \langle \bm{0}|\hat{s}_z \hat{H} |\bm{R}\rangle.
\end{equation}
The quantities need to be evaluated are the 
four real space matrices 
$\langle \bm{0}|\hat{s}_z|\bm{R}\rangle$, 
$\langle \bm{0}|\hat{s}_z (\bm{r}-\bm{R})_x|\bm{R}\rangle$, 
$\langle \bm{0}|\hat{s}_z \hat{H} (\bm{r}-\bm{R})_x|\bm{R}\rangle$ and
$\langle \bm{0}|\hat{s}_z \hat{H} |\bm{R}\rangle$. 

These real space matrices can be evaluated on the 
original coarse \textit{ab-initio} k-mesh, and we 
use the notation $|u^{(0)}_{n\bm{q}}\rangle$, 
$|u^{(H)}_{n\bm{k}}\rangle$, $|u^{(W)}_{n\bm{k}}\rangle$ and 
$|\bm{R}n\rangle$ to represent the original 
\textit{ab-initio} Bloch wave function, the Hamiltonian 
gauge wave function, the Wannier gauge wave function 
and the smooth real space Wannier wave function, respectively.
The notation $\bm{q}$ and $\bm{k}$ are used to differentiate the coarse 
\textit{ab-initio} $\bm{q}$ mesh with the dense Wannier interpolation 
$\bm{k}$ mesh.  

The four real space matrices are evaluated in the original 
\textit{ab-initio} space. The transformation from 
coarse \textit{ab-initio} q-mesh to real space mesh are 
summarized as follows,   
\begin{equation}
\label{equ:uo2uw}
|u_{n\bm{q}}^{(W)}\rangle = \sum_{m} |u_{m\bm{q}}^{(0)} \rangle V_{\bm{q},mn},
\end{equation}
\begin{equation}
\label{equ:uw2uh}
|u_{n\bm{q}}^{(H)}\rangle = \sum_m |u_{m\bm{q}}^{(W)}\rangle U_{mn},
\end{equation}
\begin{equation}
\label{equ:uw2r}
|\bm{R}n \rangle = \frac{1}{N_q^3} \sum_{\bm{q}} e^{i\bm{q}(\bm{r}-\bm{R})} |u_{n\bm{q}}^{(W)}\rangle,
\end{equation}
where $N_q^3$ is the total number of 
q-points on the coarse q-mesh.
The density of q-mesh should be sufficient that high 
quality $|\bm{R}n\rangle$ can be constructed. 

From the real space Wannier function $|\bm{R}n\rangle$, the 
real space matrices can be readily evaluated, 
\begin{equation}
\label{equ:usu_fft}
\langle \bm{0}|\hat{s}_z|\bm{R}\rangle = \frac{1}{N_q^3} 
\sum_{\bm{q}} e^{-i\bm{q}\bm{R}} \langle u_{\bm{q}}^{(W)}| \hat{s}_z | u_{\bm{q}}^{(W)} \rangle,
\end{equation}
\begin{equation}
\label{equ:usdu_fft}
\langle \bm{0}|\hat{s}_z (\bm{r}-\bm{R})_x|\bm{R}\rangle = i \frac{1}{N_q^3} 
\sum_{\bm{q}} e^{-i\bm{q}\bm{R}} \langle u_{\bm{q}}^{(W)}| \hat{s}_z | 
\partial_x u_{\bm{q}}^{(W)} \rangle,
\end{equation}
\begin{equation}
\label{equ:ushdu_fft}
\langle \bm{0}|\hat{s}_z \hat{H} (\bm{r}-\bm{R})_x|\bm{R}\rangle = i \frac{1}{N_q^3} 
\sum_{\bm{q}} e^{-i\bm{q}\bm{R}} \langle u_{\bm{q}}^{(W)}| \hat{s}_z \hat{H} | 
\partial_x u_{\bm{q}}^{(W)} \rangle,
\end{equation}
\begin{equation}
\label{equ:ushu_fft}
\langle \bm{0}|\hat{s}_z \hat{H} |\bm{R}\rangle = \frac{1}{N_q^3} 
\sum_{\bm{q}} e^{-i\bm{q}\bm{R}} \langle u_{\bm{q}}^{(W)}| \hat{s}_z \hat{H} | 
u_{\bm{q}}^{(W)} \rangle.
\end{equation}
The evaluation of the needed matrices in 
Equ.(\ref{equ:usu_fft}) and 
(\ref{equ:ushu_fft}) is fairly 
simple by using Equ.(\ref{equ:uo2uw}), 
\begin{equation}
\langle u_{\bm{q}}^{(W)}| \hat{s}_z | u_{\bm{q}}^{(W)} \rangle = 
V_{\bm{q}}^+ S_{\bm{q}}^{(0)} V_{\bm{q}},
\end{equation}
and
\begin{equation}
\langle u_{\bm{q}}^{(W)}| \hat{s}_z \hat{H} | u_{\bm{q}}^{(W)} \rangle = 
V_{\bm{q}}^+ S_{\bm{q}}^{(0)} H_{\bm{q}}^{(0)} V_{\bm{q}}.
\end{equation}
The evaluation of the needed matrices in 
Equ.(\ref{equ:usdu_fft}) and 
(\ref{equ:ushdu_fft}) utilize the smoothness of 
$| u_{\bm{q}}^{(W)} \rangle$. We expand the 
$| \partial_x u_{\bm{q}}^{(W)} \rangle$ as \cite{Marzari1997}  
\begin{equation}
| \partial_x u_{\bm{q}}^{(W)} \rangle = \sum_{\bm{b}_x} 
w_{\bm{b}_x} \bm{b}_x [u_{\bm{q}+\bm{b}_x}^{(W)} - u_{\bm{q}}^{(W)}],
\end{equation}
where $\bm{b}$ is a vector connecting a $\bm{q}$ point 
to its near neighbors and together with its weight 
$w_{\bm{b}}$ satisfy the completeness 
relation Ref. \cite{Marzari1997} Equ.(B1). Thus, 
\begin{align}
\begin{split}
& \langle u_{\bm{q}}^{(W)}| \hat{s}_z | 
\partial_x u_{\bm{q}}^{(W)} \rangle \\
= & \sum_{\bm{b}_x} w_{\bm{b}_x} \bm{b}_x 
[V_{\bm{q}}^+ \langle u_{\bm{q}}^{(0)} | \hat{s}_z | 
u_{\bm{q}+\bm{b}_x}^{(0)} \rangle V_{\bm{q}+\bm{b}_x} \\
& - V_{\bm{q}}^+ \langle u_{\bm{q}}^{(0)} | \hat{s}_z | 
u_{\bm{q}}^{(0)} \rangle V_{\bm{q}}] \\
= & \sum_{\bm{b}_x} w_{\bm{b}_x} \bm{b}_x 
[V_{\bm{q}}^+ S_{\bm{q}}^{(0)} M_{\bm{q},\bm{b}_x}^{(0)} V_{\bm{q}+\bm{b}_x} \\
& - V_{\bm{q}}^+ S_{\bm{q}}^{(0)} V_{\bm{q}}]
\end{split}
\end{align}
and 
\begin{align}
\begin{split}
& \langle u_{\bm{q}}^{(W)}| \hat{s}_z \hat{H}_{\bm{q}} | 
\partial_x u_{\bm{q}}^{(W)} \rangle \\
= & \sum_{\bm{b}_x} w_{\bm{b}_x} \bm{b}_x 
[V_{\bm{q}}^+ \langle u_{\bm{q}}^{(0)} | \hat{s}_z \hat{H}_{\bm{q}} | 
u_{\bm{q}+\bm{b}_x}^{(0)} \rangle V_{\bm{q}+\bm{b}_x} \\ 
& - V_{\bm{q}}^+ \langle u_{\bm{q}}^{(0)} | \hat{s}_z \hat{H}_{\bm{q}} | 
u_{\bm{q}}^{(0)} \rangle V_{\bm{q}}] \\
= & \sum_{\bm{b}_x} w_{\bm{b}_x} \bm{b}_x 
[V_{\bm{q}}^+ S_{\bm{q}}^{(0)} H_{\bm{q}}^{(0)} M_{q,\bm{b}_x}^{(0)} V_{q+\bm{b}_x} \\ 
& - V_{\bm{q}}^+ S_{\bm{q}}^{(0)} H_{\bm{q}}^{(0)} V_{q}]
\end{split}
\end{align}
The only unknown quantities are 
$S_{\bm{q}}^{(0)} = \langle u_{\bm{q}}^{(0)} | \hat{s}_z | 
u_{\bm{q}}^{(0)} \rangle$, 
$H_{\bm{q}}^{(0)} = \langle u_{\bm{q}}^{(0)} | \hat{H}_{\bm{q}} | 
u_{\bm{q}}^{(0)} \rangle$, 
$M_{\bm{q},\bm{b}}^{(0)} = \langle u_{\bm{q}}^{(0)} | 
u_{\bm{q}+\bm{b}}^{(0)} \rangle$ which are the spin 
operator matrices, the Hamiltonian matrices and 
the overlap matrices of the original 
\textit{ab-initio} wave function, respectively. 
Note the overlap matrices are available ``for free'' 
since they have been computed in the construction 
process of MLWF. The spin operator matrices 
and the Hamiltonian matrices are easily computed 
from \textit{ab-initio} results. 
The {\sc pw2wannier90} program, which is the 
interface between {\sc Quantum ESPRESSO} 
\cite{Giannozzi2009,Giannozzi2017Oct} and 
{\sc Wannier90} \cite{Mostofi2014}, has already calculated these three 
quantities. 
Thus by our derivations, no additional quantities 
are required for computing SHC. For other \textit{ab-initio} software 
packages, only the Hamiltonian 
matrix, which is just the eigenvalues, and the 
spin operator matrices, are the additional quantities 
that need to be passed to {\sc Wannier90} from 
\textit{ab-initio} calculation.   

At this point, all the needed matrices have been 
evaluated from the \textit{ab-initio} calculation. 
In summary, we succeed in computing the four 
real space matrices 
$\langle \bm{0}|\hat{s}_z|\bm{R}\rangle$, 
$\langle \bm{0}|\hat{s}_z (\bm{r}-\bm{R})_x|\bm{R}\rangle$, 
$\langle \bm{0}|\hat{s}_z \hat{H} (\bm{r}-\bm{R})_x|\bm{R}\rangle$ and
$\langle \bm{0}|\hat{s}_z \hat{H} |\bm{R}\rangle$. Then  
we interpolate these matrices on a dense k-mesh by 
the inverse Fourier transform, i.e. the 
Equ.(\ref{equ:usu_invfft}), 
(\ref{equ:usdu_invfft}), (\ref{equ:ushdu_invfft}) 
and (\ref{equ:ushu_invfft}). Through gauge transformation 
the $S^{(H)}$, $K^{(H)}$, $L^{(H)}$ are computed. 
Finally, through Equ.(\ref{equ:b3}) and Equ.(\ref{equ:js}) 
the spin current operator matrix is computed. 
We implemented the computer code for the Kubo formula of SHC on the basis 
of {\sc Wannier90} package \cite{Mostofi2014,Marzari1997,Souza2001}. 

\section{Computational details\label{sec:comp}}

To demonstrate the validity of our Wannier interpolation 
approach, we performed SHC calculations of fcc Pt and GaAs, 
which have been investigated by direct evaluation 
of Kubo formula Equ.(\ref{equ:kubo_shc}) based on 
LMTO method \cite{Guo2008,Guo2005}. 
After that, results of $\alpha$-Ta and $\beta$-Ta 
are shown.

All the \textit{ab-initio} calculations were performed using 
{\sc Quantum ESPRESSO} package based on projector-augmented wave (PAW) 
method and a plane wave basis set \cite{Giannozzi2009,Giannozzi2017Oct}. 
The exchange and correlation terms were described using 
generalized gradient approximation (GGA) in the scheme of 
Perdew-Burke-Ernzerhof (PBE) parameterization, as implemented in the 
{\sc pslibrary} \cite{Corso2014}. Energy convergence criteria of all the 
calculations were set as \SI{1.0e-8}{Ry}.

\subsection{\lowercase{dc} SHC \lowercase{of} \lowercase{fcc} P\lowercase{t}\label{sec:pt}}
 
For the \textit{ab-initio} calculations, we used a wave function cutoff of \SI{90}{Ry} and electron density 
cutoff of \SI{1080}{Ry}. In the self-consistent field (scf) 
calculation, Monkhorst-Pack k-meshes of $6 \times 6 \times 6$ 
to $14 \times 14 \times 14$ were tested while 
the non-self-consistent (nscf) k-meshes were kept the same as 
the scf k-meshes for the construction of 
MLWF.  
The fcc unit cell contains one Pt atom and the 
lattice constant was set as \SI{3.92}{\angstrom}. 

\begin{figure}
	\includegraphics[width=\columnwidth]{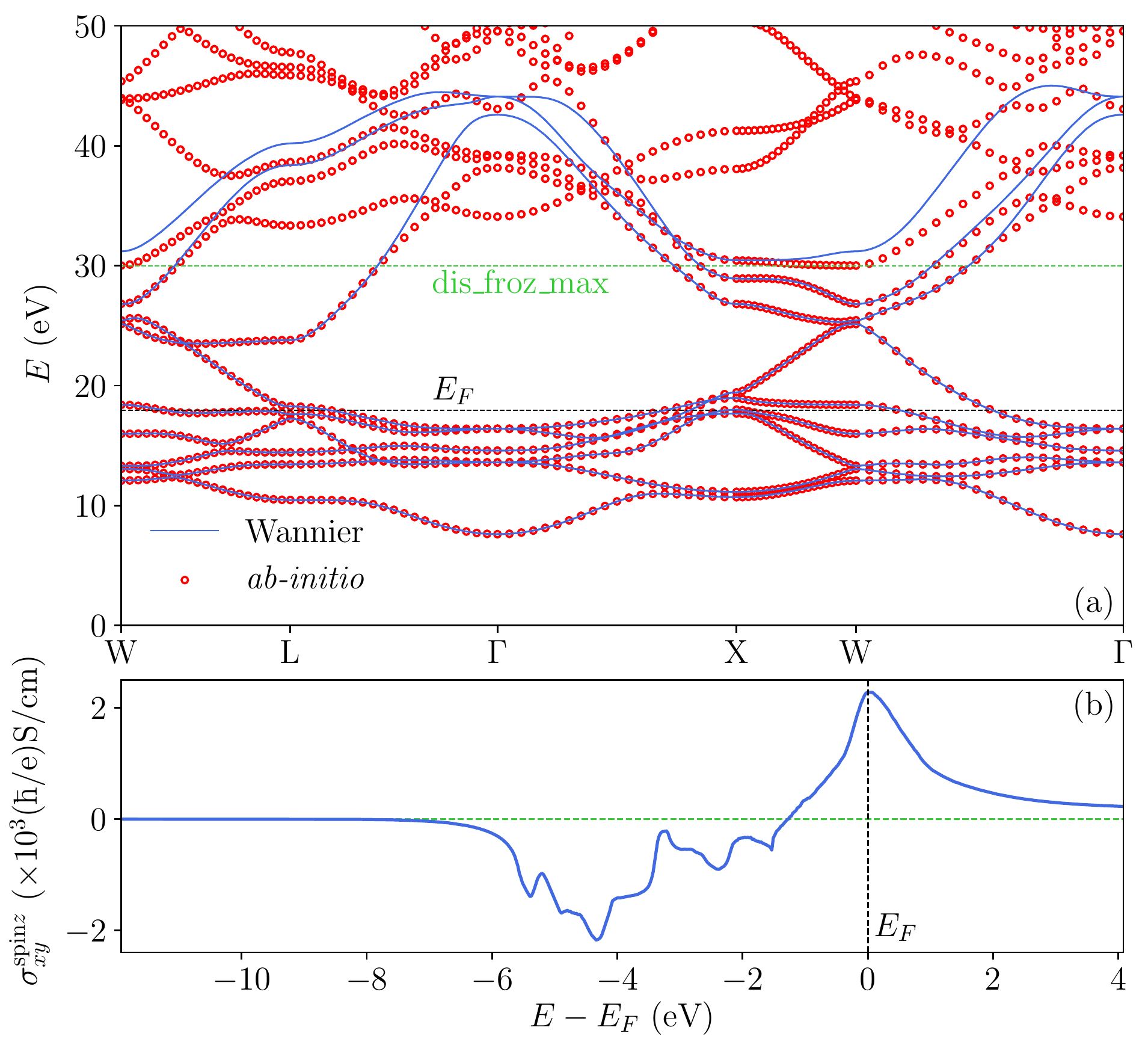}
	\caption{\label{fig:pt_band_fermiscan} 
		(a) Comparison of fcc Pt  
		band structure obtained from  
		\textit{ab-initio} calculation (red circles)		
		and the interpolated band structure 
		by MLWFs (blue lines).  
		The black dashed 
		horizontal line corresponds to the Fermi energy, 
		the green dashed horizontal line corresponds to the 
		upper limit of the frozen inner window in the 
		disentanglement process when constructing MLWFs.
		(b) The SHC variation of fcc Pt with respect to the position 
		of Fermi energy. The black dashed vertical line 
		corresponds to the calculated Fermi energy, at 
		which the SHC reaches \SI{2280}{(\hbar/e)S/cm}.}
\end{figure}

For the construction of MLWF, an inner frozen window of 
\SIrange{0.0}{30.0}{eV} and an outer disentanglement 
window of \SIrange{0.0}{60.0}{eV} were used to extract 
\num{18} spinor WFs having the form of 
$s$, $p$ and $d$-like Gaussians [see 
Fig.\ref{fig:pt_band_fermiscan}]. 
The spread of each 
WF was in the range of \SIrange{0.7}{2.3}{\angstrom^2}, 
and the spreads for both the 
disentanglement and Wannierization processes were 
converged under \SI{1e-10}{\angstrom^2}.  

As can be seen in Fig.\ref{fig:pt_band_fermiscan}, 
under the disentanglement frozen window, the \textit{ab-initio} 
band structure is fully recovered by the MLWFs. Since the Kubo 
formula Equ.(\ref{equ:kubo_shc}) involves unoccupied bands, 
a large frozen window is needed to ensure those unoccupied 
bands, which are several \si{eV} above the Fermi energy, 
can be accurately recovered. Thus the distortions of higher energy unoccupied bands have little influence on the calculated SHC. 

Comparing with the LMTO results \cite{Guo2008}, our 
result of \SI{2280}{(\hbar/e)S/cm} deviates about $3.6\%$ from that 
of \SI{2200}{(\hbar/e)S/cm}. Considering the difference 
between \textit{ab-initio} 
methods and softwares, this small deviation is tolerable and 
confirms that our Wannier interpolation approach is capable  
of calculating SHC with high accuracy. 
Varying the position 
of Fermi level can be viewed as a crude approximation of 
the doping effect on the SHC. As shown in Fig.
\ref{fig:pt_band_fermiscan}, the SHC reaches its peak 
value around the Fermi energy, while drops to 
\SI{-2172}{(\hbar/e)S/cm} at \SI{4.34}{eV} below the Fermi energy. 
The strong resemblance of the 
Fig.\ref{fig:pt_band_fermiscan}(b)  
to the result in Ref. \cite{Guo2008} again validates 
our Wannier interpolation approach. 

\begin{figure}
	\includegraphics[width=\columnwidth]{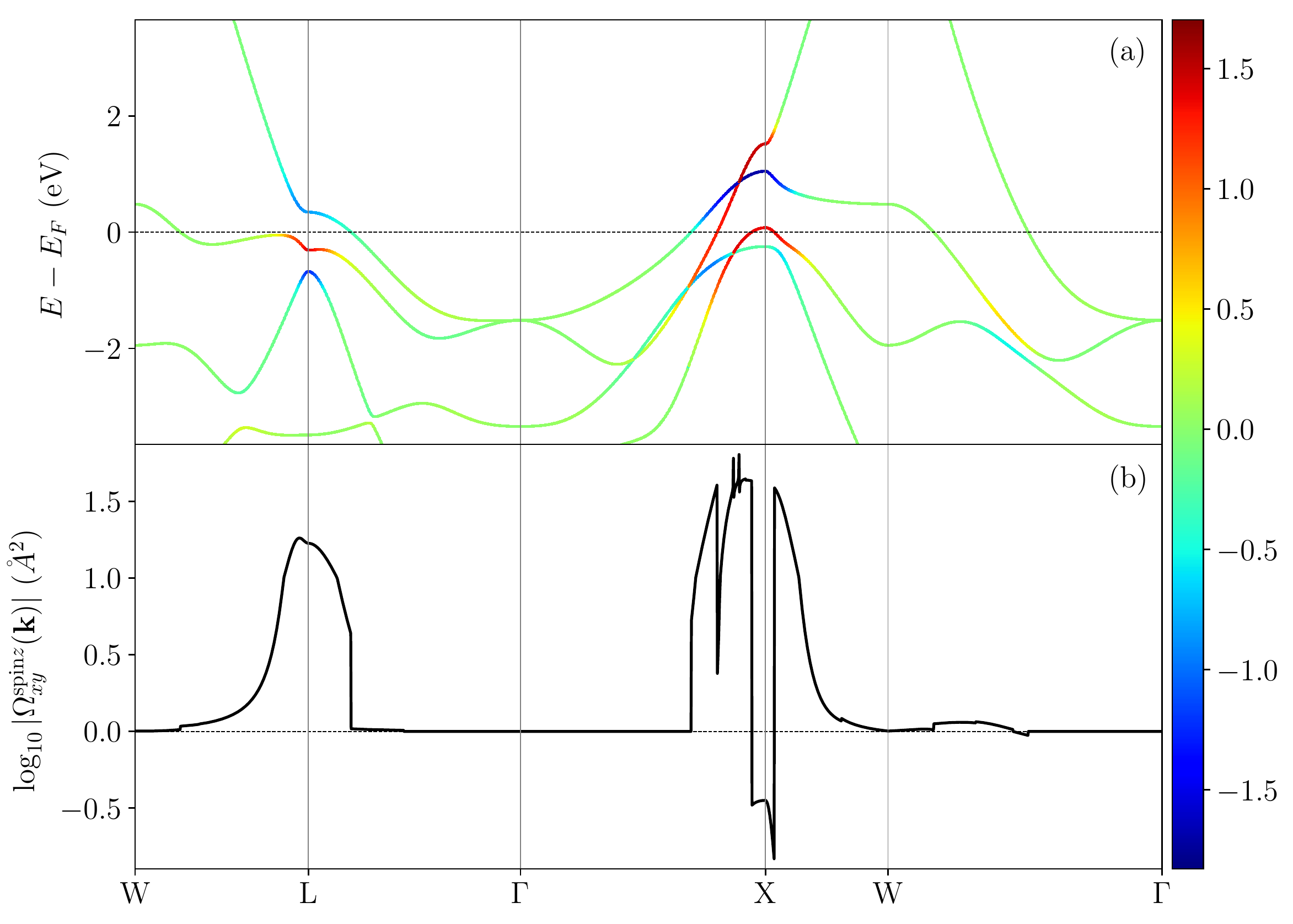}
	\caption{\label{fig:kpath} SHC of fcc Pt along a path in the BZ. 
		The color bar in the panel (a) is the SHC projected on 
		each band after taking logarithm [Equ.(\ref{equ:log})], i.e the 
		Equ.(\ref{equ:kubo_shc_berry}). The panel (b) is the 
		k-point resolved SHC after taking logarithm [Equ.(\ref{equ:log})], 
		i.e. Equ.(\ref{equ:kubo_shc_sum})
		without $\bm{k}$ sum.
	}
\end{figure}

More insights can be obtained by band projected plot of SHC, 
as shown in Fig.\ref{fig:kpath}. The color in 
Fig.\ref{fig:kpath}(a) is the SHC projected on 
each band, i.e. the magnitude  
of Equ.(\ref{equ:kubo_shc_berry}) after taking logarithm, 
\begin{equation}
\label{equ:log}
\text{result} = 
\begin{cases}
\text{sgn}(x)\cdot\log_{10}{|x|}, & |x| > 10,\\
\frac{x}{10}, & |x| \leq 10,
\end{cases}
\end{equation}
where sgn($x$) means taking the sign of $x$.
The SHC is mostly concentrated around $X$ point 
in the BZ. The small spin-orbit splitting induces large 
SHC variation. The Fig.\ref{fig:kpath}(b) is the 
k-point resolved SHC, i.e. Equ.(\ref{equ:kubo_shc_sum}) 
without $\bm{k}$ sum. Spikes near $X$ and $L$ points 
contribute to SHC significantly.

\begin{figure}
	\includegraphics[width=\columnwidth]{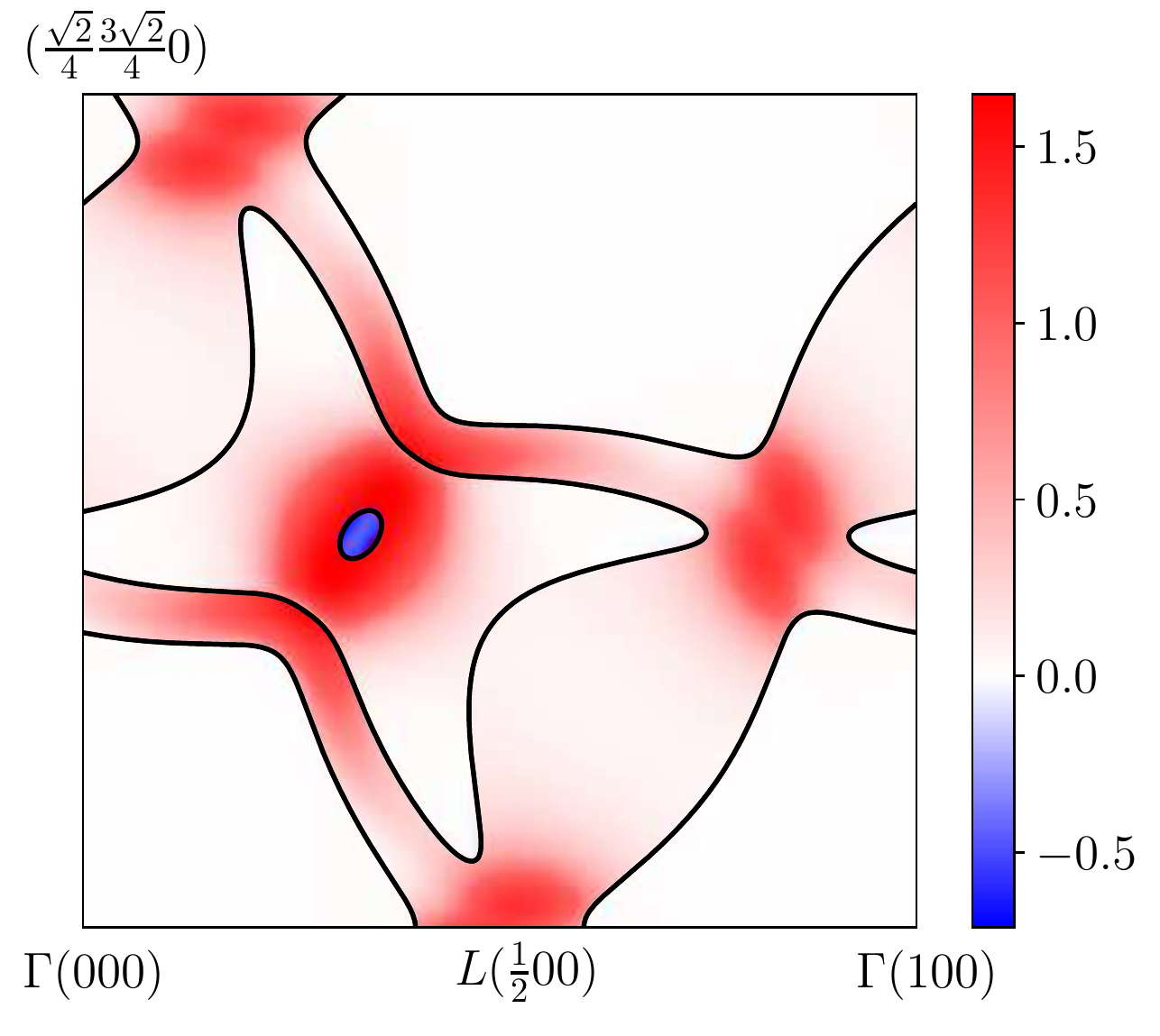}
	\caption{\label{fig:kslice} SHC of fcc Pt in a slice of 
		the BZ. The slice is perpendicular to the reciprocal space 
		$\bm{c}$ axis. The black lines are the intersection 
		of the slice with the Fermi surface. 
		The horizontal axis is the vector 
		$2\protect\overrightarrow{\Gamma L}=(1,0,0)$, 
		and the vertical axis is the vector $(\frac{\sqrt{2}}{4}, 
		\frac{3\sqrt{2}}{4},0)$, which 
		lies in the $X$-$\Gamma$-$L$ plane and is normal to 
		$\protect\overrightarrow{\Gamma L}$. The $(\frac{1}{2},0,0)$ on the 
		x-axis is the $L$ point and the near-center 
		$(\frac{1}{2},\frac{1}{2},0)$ point [i.e. 
		$(\frac{1}{3},\frac{\sqrt{2}}{3})$ relative to horizontal 
		and vertical axes] corresponds to 
		the $X$ point. The color represents the logarithm [Equ.(\ref{equ:log})] of 
		Equ.(\ref{equ:kubo_shc_sum})
		without $\bm{k}$ sum. 
		The coordinates of the above mentioned vectors are 
		fractional coordinates relative to the reciprocal lattice. }
\end{figure}

A further intuitive analysis can be carried out by plotting 
k-resolved SHC on a slice of BZ, as shown in Fig.
\ref{fig:kslice}. We chose the vector 
$2\overrightarrow{\Gamma L}=(1,0,0)$ [in fractional 
coordinates relative to reciprocal lattice] 
as the horizontal axis, and the vector $(\frac{\sqrt{2}}{4}, 
\frac{3\sqrt{2}}{4},0)$ as the vertical axis, which 
lies in the $X$-$\Gamma$-$L$ plane and is normal to 
$\overrightarrow{\Gamma L}$. The $(\frac{1}{2},0,0)$ point on the 
horizontal axis is the $L$ point, the small red 
spot around the $L$ point is consistent with the band 
projected result in Fig.\ref{fig:kpath}(a). The near-center 
$(\frac{1}{2},\frac{1}{2},0)$ point corresponds to 
the $X$ point, where rapid variation of SHC occurs. 

Since the calculation of SHC always involves k-point 
mesh on the order of million, special care must be 
taken to the convergence issues. As shown in 
the Fig.\ref{fig:kpath} and Fig.\ref{fig:kslice}, 
the rapid variation of SHC usually only occurs at a 
small portion of the full BZ. In such cases, the method 
of adaptive k-mesh refinement can be very helpful. 
The convergence of the SHC with respect to the 
choice of k-mesh is presented in Table 
\ref{tab:shc_kpt}. The value of \SI{2281.27}{(\hbar/e)S/cm} 
from the No. 11 calculation 
is regarded as the fully converged result.
Comparing the results No. 1, 2, 3 and 11, 
we find that the fully converged result can 
be conveniently obtained by adaptive k-mesh 
refinement even 
on a not so dense BZ grid. Comparing the results from No. 
4 to 9, it can be concluded that the 
Wannier interpolation k-mesh is the key parameters 
for the convergence of SHC. Besides, it is 
noticeable that the density of the original 
\textit{ab-initio} k-mesh has little 
influence on the convergence of SHC. Since in the 
construction process of MLWF, the \textit{ab-initio} 
k-mesh is sufficient if well localized real space 
quantities such as Equ.(\ref{equ:usu_fft}), 
(\ref{equ:usdu_fft}), (\ref{equ:ushdu_fft}) and 
(\ref{equ:ushu_fft}) are acquired. In fact, the 
most significant merit of MLWF is that high 
accuracy interpolated results can be obtained 
on a relatively coarse \textit{ab-initio} 
calculation, which eases the computational 
burden while preserves the 
accuracy of \textit{ab-initio} calculation. 
The above mentioned calculation of Pt, which was 
performed on $8 \times 8 \times 8$ 
\textit{ab-initio} k-mesh, $100 \times 100 \times 100$ 
Wannier interpolation k-mesh with 
$4 \times 4 \times 4$ adaptive refinement k-mesh, 
spent 492 seconds on 120 CPU cores, which is 
a bit larger than the total time spent by 
the scf and nscf calculations for the 
construction of MLWF. 

\begin{table}%
	\caption{\label{tab:shc_kpt}The SHC values and its  
		convergence trend relative to \textit{ab-initio} k-mesh, 
		Wannier interpolation k-mesh, and adaptive refinement 
		k-mesh. All the k-mesh numbers $n$ in each cell of 
		the table should be expanded as $n \times n \times n$. 
		For example the No. 1 corresponds to the 
		calculation performed on $6 \times 6 \times 6$ 
		\textit{ab-initio} k-mesh and $100 \times 100 \times 100$ 
		Wannier interpolation k-mesh without adaptive k-mesh 
		refinement.}
	\begin{ruledtabular}
		\begin{tabular}{c c c c c}
			No. & ab. k-mesh & Wan. k-mesh & adpt. k-mesh & SHC(\si{(\frac{\hbar}{e})S/cm}) \\
			\hline
			\num{1} & \num{6} & \num{100} & \num{0} & \num{2559.39} \\
			\num{2} & \num{6} & \num{100} & \num{2} & \num{2281.44} \\
			\num{3} & \num{6} & \num{100} & \num{4} & \num{2281.44} \\
			\hline
			\num{4} & \num{8} & \num{40} & \num{0} & \num{2300.56} \\
			\num{5} & \num{8} & \num{60} & \num{0} & \num{2270.44} \\
			\num{6} & \num{8} & \num{80} & \num{0} & \num{2286.38} \\
			\num{7} & \num{8} & \num{100} & \num{0} & \num{2279.62} \\
			\num{8} & \num{8} & \num{120} & \num{0} & \num{2282.64} \\
			\num{9} & \num{8} & \num{140} & \num{0} & \num{2283.84} \\
			\hline
			\num{10} & \num{14} & \num{200} & \num{0} & \num{2282.30} \\
			\num{11} & \num{14} & \num{300} & \num{0} & \num{2281.27} \\
		\end{tabular}
	\end{ruledtabular}
\end{table}

\subsection{\lowercase{ac} SHC \lowercase{of} G\lowercase{a}A\lowercase{s}\label{sec:gaas}}

We further performed ac SHC calculations 
of GaAs. 
In this case, the Monkhorst-Pack k-meshes of 
$10 \times 10 \times 10$ were used 
for the scf and nscf calculations. 
The fcc unit cell contains one Ga atom located 
at $(0,0,0)$ and one As atom located at 
$(\frac{1}{4},\frac{1}{4},\frac{1}{4})$, and the 
lattice constant was set as \SI{5.654}{\angstrom}. 
The \textit{ab-initio} and the Wannier 
interpolated band structures are shown in 
supplemental material \cite{sm} Fig.
\ref{fig:gaas_band}. The calculated band gap 
of \SI{0.40}{eV} at $\Gamma$ point is much 
smaller than the experimental value of 
\SI{1.52}{eV}, so we applied a scissors shift 
of the conduction band so that the experimental  
band gap can be restored. 

For the construction of MLWF, the 10 underlying 
valence bands are excluded, and 16 WFs are 
extracted from the 16 bands around the band 
gap. No disentanglement process is adopted 
since the 16 bands are isolated.  
\num{16} spinor WFs having the form of 
As located $sp^3$-like and Ga located $sp^3$-like 
Gaussians are used [see supplemental material \cite{sm}
Fig.\ref{fig:gaas_band}]. 
The spread of each 
WF is in the range of \SIrange{3.6}{4.7}{\angstrom^2}, 
and the spread for the Wannierization process was 
converged under \SI{1e-10}{\angstrom^2}. 
The Wannier interpolated band structure again 
accurately recovered the \textit{ab-initio} one.

\begin{figure}
	\includegraphics[width=\columnwidth]{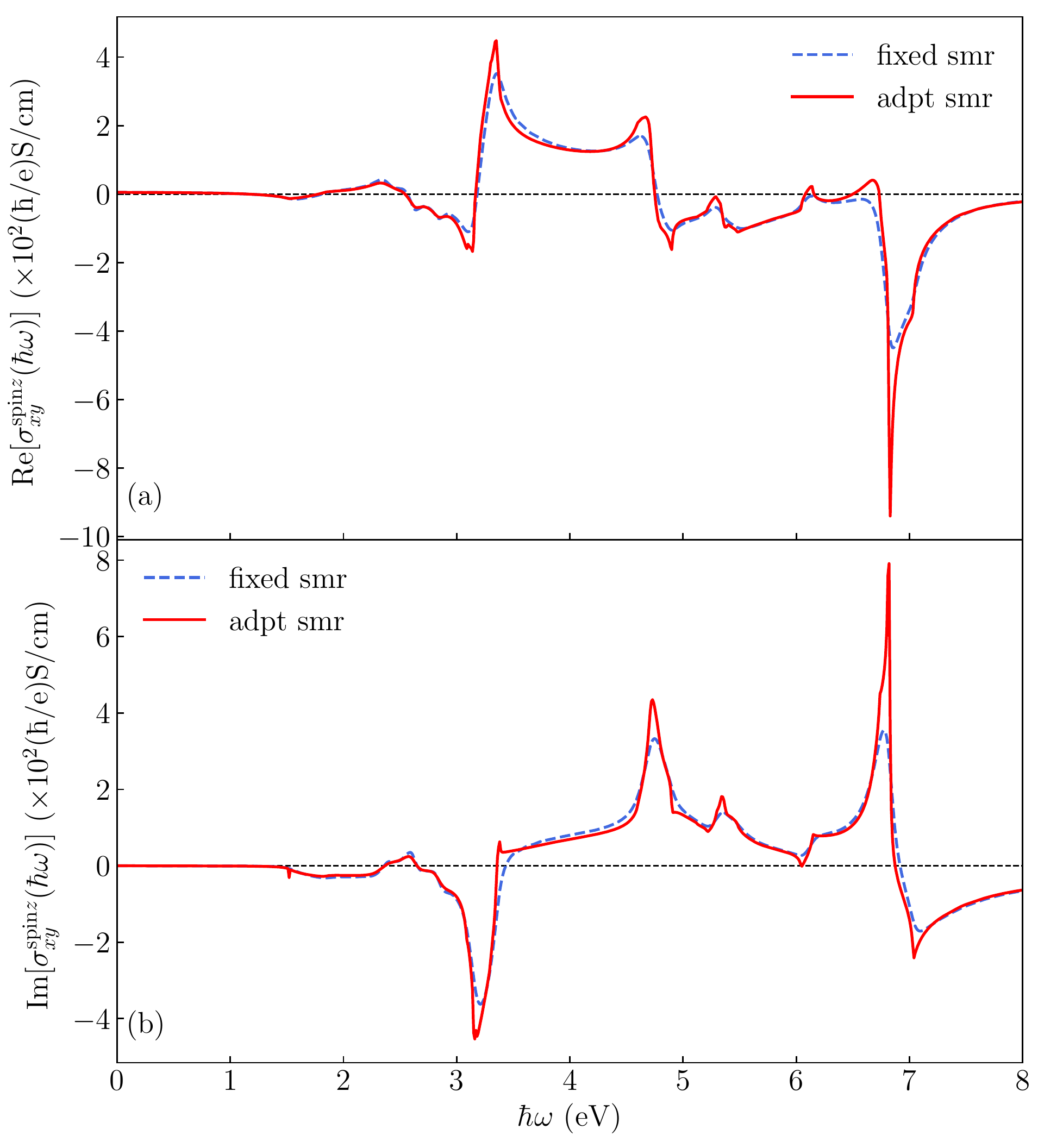}
	\caption{\label{fig:gaas_freqscan} The (a) real part and 
		(b) imaginary part of GaAs ac SHC. The blue dashed 
		lines correspond to calculation with fixed smearing width 
		of \SI{0.05}{eV}, while the red solid lines correspond to calculation 
		with adaptive smearing.}
\end{figure}

When the frequency $\omega$ varies, the denominator in 
Equ.(\ref{equ:kubo_shc}) may approach 0, causing large 
spikes in the plot. To mitigate this difficulty, two 
kinds of smearing are adopted, the fixed smearing and 
adaptive smearing. Follow the method in Ref. \cite{Yates2007}, 
the smearing parameter $\eta$ is set as 
\begin{equation}
\label{equ:adpt_smr}
W_{nm,\bm{k}} = a\left|\frac{\partial\mathcal{E}_{n\bm{k}}}
{\partial\bm{k}}-\frac{\partial\mathcal{E}_{m\bm{k}}}
{\partial\bm{k}}\right|\Delta k,
\end{equation}
which is varied according to the level spacing. The $a$ is 
the factor of adaptive smearing, and further more 
the calculated $W_{nm,\bm{k}}$ is compared with a fixed 
value $W_{max}$ to avoid too large smearing parameter. In the 
calculation of Fig.\ref{fig:gaas_freqscan}, we set 
$a=1.414$ and $W_{max}=\SI{1.0}{eV}$, while for fixed smearing 
we set $W=\SI{0.05}{eV}$. 
As shown in Fig.\ref{fig:gaas_freqscan}, the adaptive 
smearing method successfully avoids unphysical reduction 
of the peaks and kinks, thus the characteristics of Van Hove 
singularities are well preserved. 

Irrespective of the different \textit{ab-initio} 
method adopted in the calculations, our results of 
Fig.\ref{fig:gaas_freqscan} still well match that 
of Ref. \cite{Guo2005}, further validating the 
accuracy of our Wannier interpolation approach.

\subsection{$\alpha$-Ta and $\beta$-Ta\label{sec:ta}}

In this section we perform SHC calculations on tantalum. 
Tantalum has been adopted in many SOT experiments because of 
its large spin Hall angle (SHA). Two phases of tantalum can 
exist, i.e. the $\alpha$-Ta and $\beta$-Ta. It is already 
known that the resistivity of $\beta$-Ta, around 
\SI{200}{\mu\Omega\cm}, is approximately 4 times as large as
that of $\alpha$-Ta, which has resistivity around 
\SI{50}{\mu\Omega\cm} \cite{Read1965Aug,Clevenger1992Nov}. 
This partly explains the relatively large SHA of $\beta$-Ta, 
since SHA is proportional to the 
ratio of SHC $\sigma_{xy}^{\text{spin}z}$ to conductivity 
$\sigma_{xx}$. For a further understanding of Ta SHA, a 
comparison on SHC of $\alpha$-Ta and $\beta$-Ta is beneficial.

For the \textit{ab-initio} calculations, we used a wave function cutoff of \SI{60}{Ry} and electron density 
cutoff of \SI{480}{Ry}. For the self-consistent field (scf) 
calculation, Monkhorst-Pack k-mesh of $8 \times 8 \times 8$ 
was used, and 
the non-self-consistent (nscf) k-mesh was kept the same as 
the scf k-mesh for the construction of 
MLWF. The $\alpha$-Ta 
has a bcc crystal structure and the lattice constant was set as the relaxed result of \SI{3.322}{\angstrom}. 
For the Wannier interpolation, a k-mesh of 
$100 \times 100 \times 100$ and an adaptive k-mesh of 
$10 \times 10 \times 10$ were used. For the calculations of band projected 
and k-resolved SHC, i.e. Fig.\ref{fig:alpha_kpath}, 
\ref{fig:beta_kpath}, \ref{fig:alpha_kslice} and 
\ref{fig:beta_kslice} 
[in supplemental material \cite{sm}], 
a fixed smearing width 
of \SI{0.05}{eV} were adopted. 

For the construction of MLWF, an inner frozen window of 
\SIrange{0.0}{30.0}{eV} and an outer disentanglement 
window of \SIrange{0.0}{50.0}{eV} were used to extract 
\num{18} spinor WFs having the form of 
$d$, $p$ and $s$-like Gaussians [see 
supplemental material \cite{sm}
Fig.\ref{fig:alpha_band}]. 
The spread of each 
WF was less than \SI{1.64}{\angstrom^2}, 
and the spreads for both the 
disentanglement and Wannierization processes were 
converged under \SI{1e-10}{\angstrom^2}.  
As can be seen in supplemental material \cite{sm} Fig.\ref{fig:alpha_band}, 
under the disentanglement frozen window, the \textit{ab-initio} 
band structure is again fully recovered by the MLWFs.

\begin{figure}[htb]
	\includegraphics[width=\columnwidth]{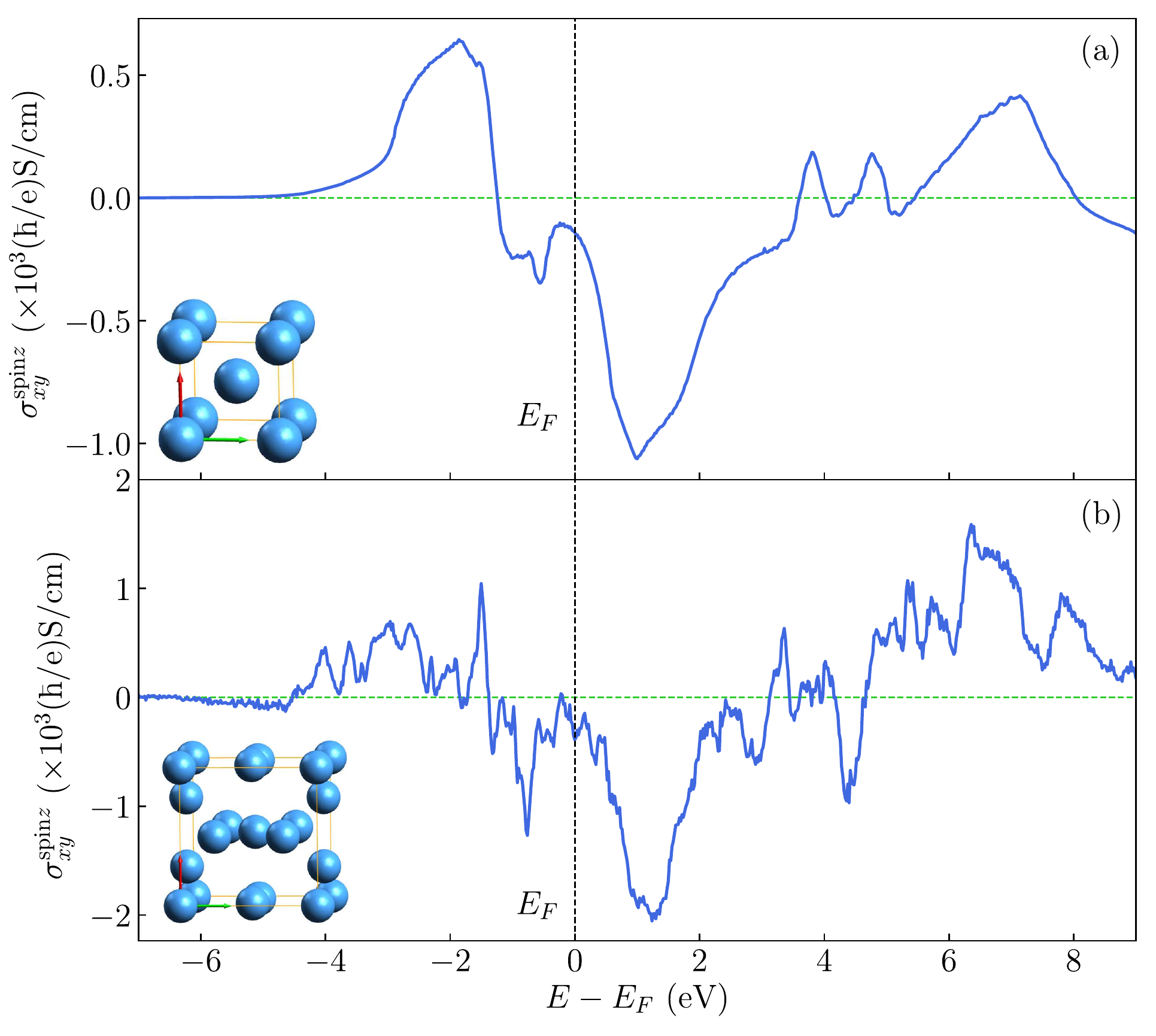}
	\caption{\label{fig:alpha_beta_fermiscan_struct} 
		The variation of SHC with respect to the position 
		of Fermi energy for (a) $\alpha$-Ta and 
		(b) $\beta$-Ta. The black dashed vertical line 
		corresponds to the calculated Fermi energy, at 
		which the SHC reaches \SI{-142}{(\hbar/e)S/cm} for 
		$\alpha$-Ta and \SI{-389}{(\hbar/e)S/cm} for $\beta$-Ta.
	The inset 
	shows the crystal structures of $\alpha$-Ta: bcc
	and $\beta$-Ta: tetragonal.}
\end{figure}

The variation of SHC with respect to the changes 
of Fermi energy is shown in Fig.
\ref{fig:alpha_beta_fermiscan_struct}(a). The SHC at Fermi energy 
is \SI{-142}{(\hbar/e)S/cm}, which is an order of magnitude smaller 
than that of fcc Pt. While at \SI{1}{eV} above 
Fermi energy, the SHC reaches its peak 
value of 
\SI{-1062}{(\hbar/e)S/cm}. 
This can be further comprehended by 
band projected and k-point resolved SHC, 
as shown in Fig.\ref{fig:alpha_kpath}. 
The contribution to SHC is mostly concentrated around $P$ point and along $\Gamma$ to $H$ path in the BZ. Band crossings and small spin-orbit-split gaps induce large variations of
SHC. When Fermi energy are raised \SI{1}{eV} higher, 
the contributions of $P$ point which are located 
above the Fermi energy, will be included in the 
sum of Equ.(\ref{equ:kubo_shc_sum}). When 
Fermi energy is raised even higher, contributions 
with positive sign [red color in Fig.\ref{fig:alpha_kpath}(a)] will cancel the 
aforementioned minus sign contributions [blue color 
in Fig.\ref{fig:alpha_kpath}(a)]. Thus the peaks 
appear and then disappear in 
Fig.\ref{fig:alpha_beta_fermiscan_struct}(a).
The k-resolved SHC on a slice of BZ, 
as shown in supplemental material \cite{sm} Fig.
\ref{fig:alpha_kslice}, shows similar 
characteristics as that of Pt Fig.\ref{fig:kslice}.
Rapid variations of SHC occurs near Fermi 
surface, where spin-orbit coupling induces 
avoided crossings. 

\begin{figure}[htb]
	\includegraphics[width=\columnwidth]{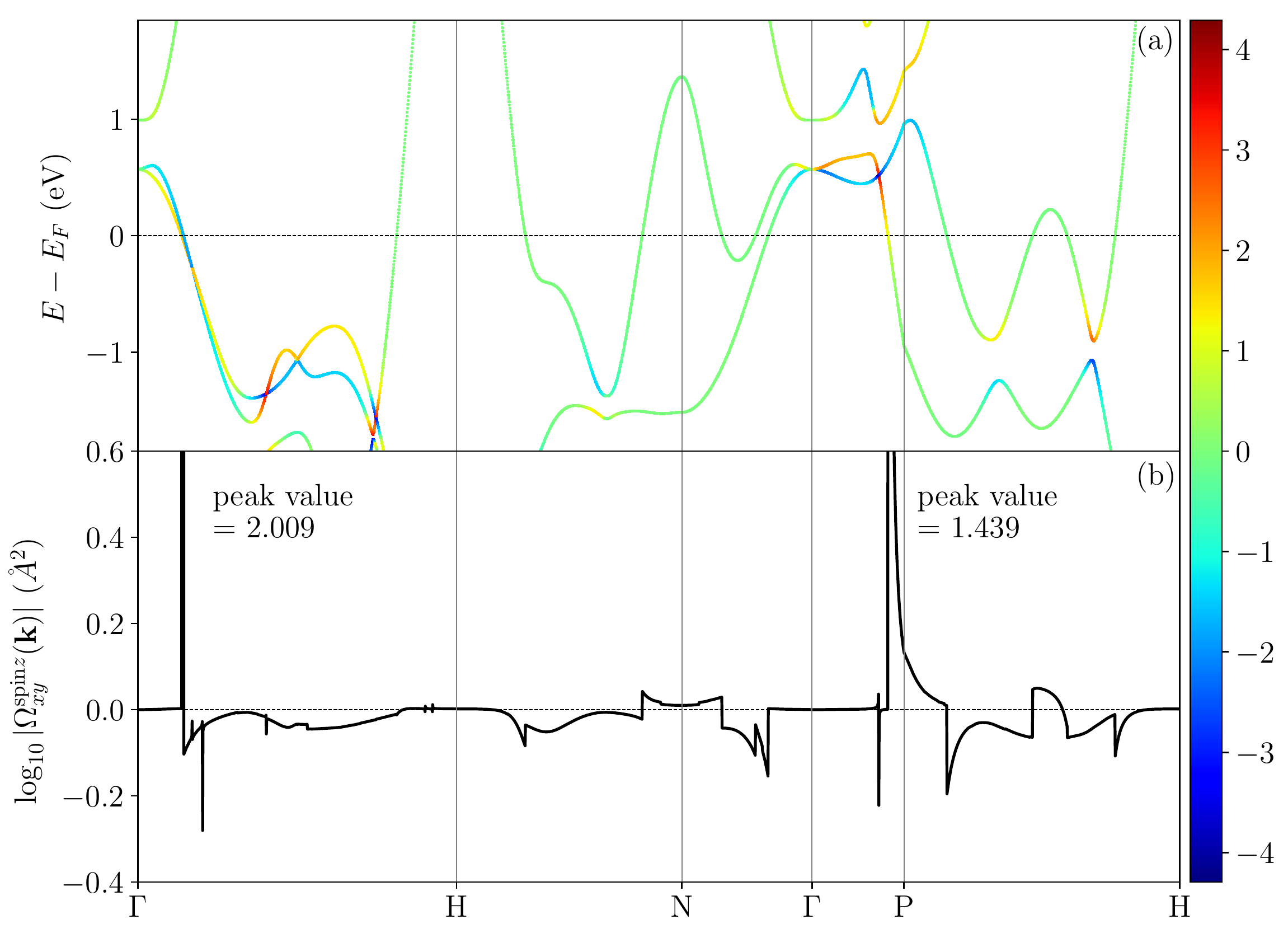}
	\caption{\label{fig:alpha_kpath} SHC of 
		$\alpha$-Ta along a path in the BZ. 
		The color bar in the panel (a) is the SHC projected on 
		each band after taking logarithm [Equ.(\ref{equ:log})], i.e the 
		Equ.(\ref{equ:kubo_shc_berry}). The panel (b) is the 
		k-point resolved SHC after taking logarithm [Equ.(\ref{equ:log})], 
		i.e. Equ.(\ref{equ:kubo_shc_sum}) 
		without $\bm{k}$ sum.
	}
\end{figure}


We further performed SHC calculation 
of $\beta$-Ta. The case of $\beta$-Ta is more 
complicated than $\alpha$-Ta. Regarding the crystal 
structure of $\beta$-Ta, models of tetragonal, 
hexagonal, $\beta$-Uranium and etc. have been 
proposed \cite{Jiang2003Aug}. Some recent 
experiments find $\beta$-Ta has tetragonal 
structure \cite{Hahn2013May,Yu2018Jul}, thus
we adopt tetragonal structure in the SHC 
calculation of $\beta$-Ta, as shown in the 
inset of Fig.\ref{fig:alpha_beta_fermiscan_struct}(b). 

The unit cell contains 8 Ta atoms. The lattice 
constants were set as $a=$\SI{5.34}{\angstrom} 
and $c$=\SI{4.97}{\angstrom}. 
For the construction of MLWF, an inner frozen window 
of \SIrange{0.0}{30.0}{eV} and an outer disentanglement 
window of \SIrange{0.0}{50.0}{eV} were used to extract 
\num{144} spinor WFs having the form of 
$d$, $p$ and $s$-like Gaussians for each Ta atom. 
A Monkhorst-Pack k-mesh of 
$4 \times 4 \times 4$ was used 
in the nscf calculation and it 
was found that the spread of each 
WF was less than \SI{1.86}{\angstrom^2}, 
and the spreads for both the 
disentanglement and Wannierization processes were 
converged under \SI{1e-10}{\angstrom^2}.   
For the Wannier interpolation, a k-mesh of 
$60 \times 60 \times 60$ and an adaptive k-mesh of 
$6 \times 6 \times 6$ were used.
The Wannier interpolated band structure again 
accurately recovered the \textit{ab-initio} one, 
shown in supplemental material 
Fig.\ref{fig:beta_band}. 

Unlike the case of $\alpha$-Ta, the variation 
of SHC relative to the position of Fermi energy 
shows more complex behavior. The SHC at Fermi energy 
is \SI{-389}{(\hbar/e)S/cm}, still an order of magnitude smaller 
than that of fcc Pt, but 2.7 times of 
$\alpha$-Ta. While at \SI{1.238}{eV} above 
Fermi energy, the SHC reaches its peak 
value of 
\SI{-2055}{(\hbar/e)S/cm}. The bands around 
Fermi energy are mainly composed of $d$ states, 
where considerable numbers of band crossings and
spin-orbit-split gaps induce large changes of 
SHC, consistent with the band projected SHC as 
shown in Fig.\ref{fig:beta_kpath}(a). 
k-resolved SHC on a slice of BZ is shown in 
supplemental material \cite{sm} Fig.
\ref{fig:beta_kslice}. 
Same as before, rapid variations 
of SHC happen again in the vicinity of Fermi 
surface. 

\begin{figure}[htb]
	\includegraphics[width=\columnwidth]{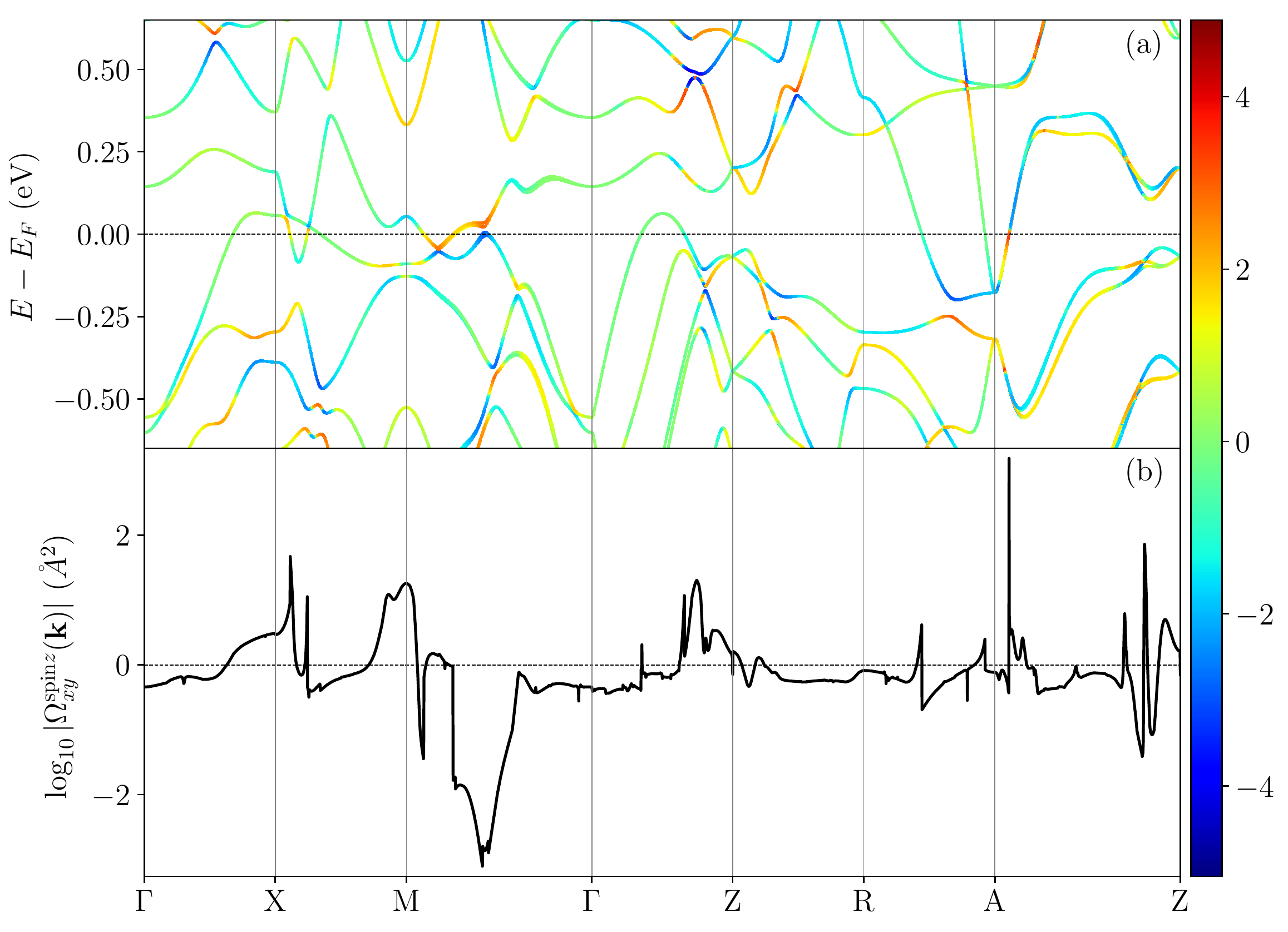}
	\caption{\label{fig:beta_kpath} SHC of $\beta$-Ta along a path in the BZ. 
		The color bar in the panel (a) is the SHC projected on 
		each band after taking logarithm [Equ.(\ref{equ:log})], i.e the 
		Equ.(\ref{equ:kubo_shc_berry}). The panel (b) is the 
		k-point resolved SHC after taking logarithm [Equ.(\ref{equ:log})], 
		i.e. Equ.(\ref{equ:kubo_shc_sum}) 
		without $\bm{k}$ sum.}
\end{figure}

In summary, SHC of $\beta$-Ta is \SI{-389}{(\hbar/e)S/cm}, while 
that of $\alpha$-Ta is \SI{-142}{(\hbar/e)S/cm}. The SHC of 
$\beta$-Ta is 2.7 times of $\alpha$-Ta, combined with 
the larger resistivity of $\beta$-Ta, resulting in the 
larger SHA. Based on experimental results 
of resistivity $\rho_{xx}^{\alpha\text{-Ta}} \simeq 
$\SI{50}{\mu\Omega\cm} and $\rho_{xx}^{\beta\text{-Ta}} \simeq $\SI{200}{\mu\Omega\cm}, we can evaluate 
the SHA according to 
\begin{equation}
\theta_{SH} = \frac{2e}{\hbar} \frac{\sigma_{xy}}{\sigma_{xx}}.
\end{equation}
The results are 
$\theta_{SH}^{\alpha\text{-Ta}} \simeq -0.0142$ and 
$\theta_{SH}^{\beta\text{-Ta}} \simeq -0.156$. 
The magnitude of $\theta_{SH}^{\beta\text{-Ta}}$ is 
quite consistent with the experimental value of 
\numrange{0.12}{0.15} \cite{Liu2012}. 
Considering the resistivity of Ta is located 
near the good metal regime of $\sigma_{xx} \simeq 10^4
\text{-}10^6$\si{S/cm} \cite{Sinova2015,Nagaosa2010May}, 
the intrinsic contribution 
should dominate so the experimental result can be 
readily reproduced by our calculation.  
Moreover, the signs of SHC for 
both $\alpha$-Ta and $\beta$-Ta are opposite to that of 
fcc Pt, and this sign difference has been verified by 
experiment \cite{Liu2012,Yu2018Jul}. In addition, it is worth 
mentioning that the SHC of both $\alpha$-Ta and 
$\beta$-Ta reach its peak at around \SI{1}{eV} above 
their respective Fermi energy, doping or alloying with 
other materials may shift the Fermi energy to reach 
the maximum SHC.

\section{Summary\label{sec:sum}}

The Wannier interpolation approach for the Kubo formula 
of intrinsic SHC is developed to achieve 
high accuracy and efficiency. The results 
of Pt dc SHC and GaAs ac SHC are 
validated against previous works by direct evaluating 
Kubo formula, and SHC of $\alpha$-Ta and 
$\beta$-Ta are calculated based on the Wannier 
interpolation approach. It is found 
that SHC of $\beta$-Ta is 2.7 times of 
$\alpha$-Ta, while both have the opposite sign relative 
to fcc Pt and are an order of magnitude smaller than Pt. 
Moreover, based on experimental data of 
resistivity, our calculated spin Hall angle of 
$\beta$-Ta is -0.156, quite consistent with 
spin Hall angle measured in experiment. This 
further implies that intrinsic contribution dominates 
in the spin Hall effect of $\beta$-Ta. 
 
The calculations are performed in 
four consecutive steps. First, a self-consistent calculation 
produces the converged charge densities. Second, a non-self-consistent 
calculation is performed on a regular k-mesh and 
three matrices are computed, which are the spin operator 
matrix, the overlap matrix and the Hamiltonian matrix. 
Third, the maximally localized Wannier functions are 
constructed. Finally, the Kubo formula is interpolated 
on a dense k-mesh by MLWF to obtain converged results. 
Due to the merit of real space localization, the fourth 
interpolation step is very efficient compared with 
``brute-force'' \textit{ab-initio} calculation on 
the dense k-mesh. 

To facilitate the convergence of SHC, adaptive refinement 
of k-mesh is implemented. 
The rapid variations of SHC, which 
are usually located in small portion of the full Brillouin 
zone, can be captured by the adaptive k-mesh refinement 
effectively. Since calculations based on 
LDA/GGA exchange correlation potentials 
often predict smaller band gaps than experiments, we implement 
scissors shift in the calculation of SHC to 
rectify this deviation. 
To improve convergence in ac SHC calculations, 
adaptive smearing is implemented and the GaAs calculation 
shows desirable results.

This Wannier interpolation approach serves as 
a post-processing step to economically calculate SHC. 
The ultimate accuracy of the calculated SHC is determined 
by the underlying \textit{ab-initio} code, 
since the MLWFs are constructed to provide lossless 
interpolation of physical quantities in the energy range 
of interest. The construction of MLWF is independent from the 
choice of the underlying code, thus our derivation and 
implementation of Wannier interpolation for SHC are able 
to co-operate seamlessly with different \textit{ab-initio} 
algorithms and implementations.

\begin{acknowledgments}

The authors gratefully acknowledge the National Natural Science 
Foundation of China (Grant No. 61627813, 61571023), the International 
Collaboration Project B16001, and the National Key Technology Program 
of China 2017ZX01032101 for their financial support of this work.

\end{acknowledgments}

%

\end{document}